\documentclass[twocolumn,10pt,showpacs,pra,superscriptaddress]{revtex4-1}
\usepackage{graphicx}
\usepackage{amsfonts}
\usepackage{amsmath}
\usepackage{siunitx}
\usepackage{hyperref}

\begin{document}
\renewcommand{\figurename}{Figure}
\sisetup{range-phrase=--}
\sisetup{range-units=single}

%
%
\title{Predicting the motion of a high-Q pendulum subject to seismic perturbations\\using machine learning}

\author{Nicolas Heimann}
\email{nicolas.heimann@uni-hamburg.de}
\affiliation{Zentrum f\"ur Optische Quantentechnologien and Institut f\"ur Laserphysik, Universit\"at Hamburg, 22761 Hamburg, Germany}
\affiliation{The Hamburg Centre for Ultrafast Imaging, Luruper Chaussee 149, 22761 Hamburg, Germany}
\author{Jan Petermann}
\author{Daniel Hartwig}
\author{Roman Schnabel}
\affiliation{Zentrum f\"ur Optische Quantentechnologien and Institut f\"ur Laserphysik, Universit\"at Hamburg, 22761 Hamburg, Germany}
\author{Ludwig Mathey}
\affiliation{Zentrum f\"ur Optische Quantentechnologien and Institut f\"ur Laserphysik, Universit\"at Hamburg, 22761 Hamburg, Germany}
\affiliation{The Hamburg Centre for Ultrafast Imaging, Luruper Chaussee 149, 22761 Hamburg, Germany}

\begin{abstract}
	The seismically excited motion of high-Q pendula in gravitational-wave observatories sets a sensitivity limit to sub-audio gravitational-wave frequencies.
	Here, we report on the use of machine learning to predict the motion of a high-Q pendulum with a resonance frequency of \SI{1.4}{Hz} that is driven by natural seismic activity.
	We achieve a reduction of the displacement power spectral density of \SI{40}{dB} at the resonant frequency \SI{1.4}{Hz} and \SI{6}{dB} at \SI{11}{Hz}.
	Our result suggests that machine learning is able to significantly reduce seismically induced test mass motion in gravitational-wave detectors in combination with corrective feed-forward techniques. 
\end{abstract}

\maketitle

%
%
\section{Introduction}
\begin{figure*}
	\includegraphics{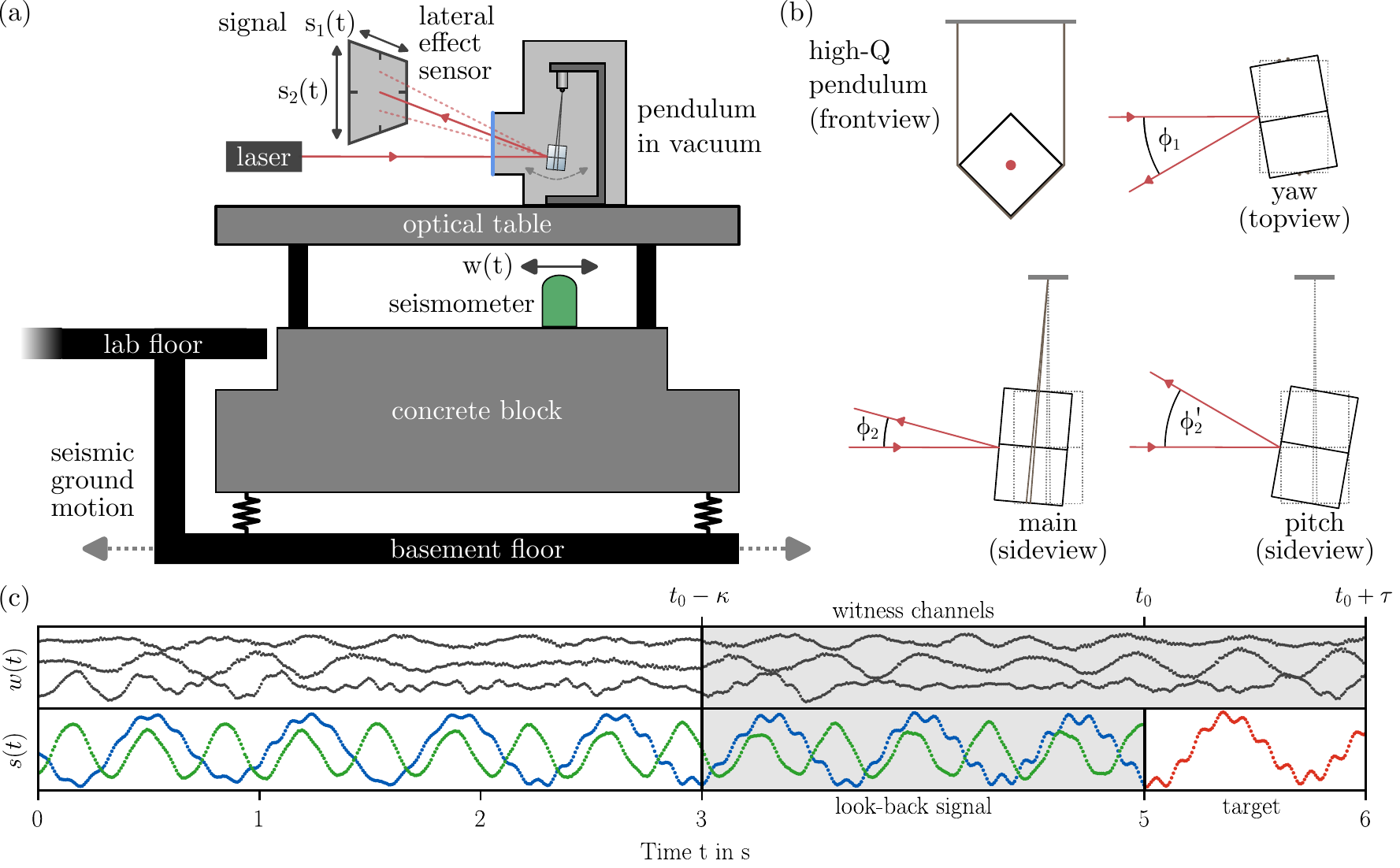}
	\caption{Sketch of the experiment and the machine learning method.
	\textbf{(a)} The interferometer is placed on top of an optical table decoupled from the environment by a 30 metric tons suspended concrete platform.
	\textbf{(b)} Illustration of the relevant modes of the suspended test mass. 
	The dotted lines show the pendulum in its equilibrium position.
	In the main pendulum mode, the entire pendulum rotates around the suspension point, such that the center of mass is moving.
	For the pitch and yaw mode, the test mass rotates around its center of gravity.
	The deflection angles $\phi_1$ and $(\phi_2 + \phi_2')$ of the laser beams are measured by a lateral effect sensor.
	\textbf{(c)} The basic construction of the method invokes look-back signal $s_1(t)$ (green), $s_2(t)$ (blue) from the photodetector and seismic witness channels $w(t)$ (grey) to forecast the target signal $y(t)$ (red).}
	\label{fig:sketch}
\end{figure*}

Pendulum suspensions are used to isolate sensitive experiments from seismic and other environmental disturbances~\cite{Robertson2002,dattilo_virgo_2003,Li2018}.
The inertia of the pendulum mass resists the motion of the suspension point at frequencies higher than the pendulum resonance.
At resonant frequencies, the movement of the suspension point is amplified.
This behavior applies to pendulum-suspended mirrors and is used to passively stabilize the optical path length in high-precision interferometry at audio-band frequencies, such as gravitational-wave astronomy~\cite{Aasi_2015,Matichard_2015}.
Seismic ground motion couples to the pendulum motion in two ways, via the mechanical contact and via the gravitational force due to fluctuations of the gravitational field known as Newtonian noise ~\cite{driggers_subtraction_2012}.
This presents a major challenge for the next generation gravitational-wave detectors at sub audio-band signal frequencies as Newtonian noise can not be shielded~\cite{harms_lower_2022} and has to be reduced by other strategies.
Another source of displacement fluctuations in pendulum suspensions is thermal noise, imposing a high quality (high-Q) factor requirement on the pendulum mode~\cite{Gonzalez2000}.

Machine learning is a broad and versatile framework for data interpretation and task optimization.
Given the data intense operation of gravitational-wave detectors, and the significant need to optimize measurements at a high precision, machine learning is a natural toolbox to utilize~\cite{Cuoco2021}.
Recent applications include noise subtraction~\cite{Vajente_2020,Ormiston_2020}, and the classification of transients~\cite{razzano_image-based_2018}.
Further applications are gravitational waveform modeling~\cite{doctor_statistical_2017}, gravitational-wave signal searches~\cite{baker_multivariate_2015}, astrophysical interpretation of gravitational-wave sources~\cite{graff_bambi_2012} and optimization of sensor placement for Newtonian noise cancellation~\cite{badaracco_machine_2020}.
Machine learning is hence a growing technology in gravitational-wave astronomy already serving fruitful results over a wide spectrum of challenges.

In this work we present a machine learning based multivariate time-series forecasting model aided by witnessed seismic noise.
We construct a high-Q factor pendulum on a passive isolation platform subject to environmental noise.
Our model allows to reconstruct the motion of the pendulum at frequencies below \SI{25}{\hertz} and we show that utilizing witnessed seismic noise from a seismometer enhances the predictive capabilities by over a magnitude.
We argue that machine learning based active stabilization offers a promising platform to enhance the signal to noise ratio in next generation gravitational-wave detectors.

This paper is organized as follows.
In Section II we describe the high-Q factor pendulum setup on top of the passive stabilization platform.
In Section III, we present two models, one without access to seismic witness channels (model~I) and a second one (model~II) utilizing witnessed seismic noise from a seismometer located on the passive stabilization platform.
In Section IV we compare the predictive capabilities of both models.
In Section V we conclude our findings.
In Section VI we provide an outlook regarding corrective feed-forward techniques.

%
%
\section{High-Q factor pendulum setup}
At the heart of our setup is a fused silica test mass, suspended as a pendulum inside a high vacuum environment with a pendulum mode resonance frequency of $f_0=\SI{1.435}{\hertz}$ and a Q factor of $Q_0 = 6 \cdot 10^4$.
The vacuum mitigates the damping due to friction of the test mass with the surrounding gas as well as coupling of acoustic disturbances.
The vacuum chamber is mounted on an optical table located on a passive seismic isolation platform that extends into the basement floor.
This platform is composed of a concrete block with a mass of approximately 30 metric tons suspended on helical springs.
In Figure~\ref{fig:sketch}~(a), we show an illustration of this setup.
Near the resonant frequencies of the platform~$f_s^{x,y,z}$, ground motion is amplified which adds to the excitation of the pendulum modes.
The relevant degrees of motion of the test mass are the main-, pitch- and yaw-mode as illustrated in Figure~\ref{fig:sketch}~(b).

We measure the deflection angles~$\phi_1$ and $(\phi_2 + \phi_2')$ of a \SI{1064}{\nano\meter} laser beam reflected off one surface of the test mass.
This measurement is performed by guiding the reflected light to a lateral effect sensor (Thorlabs PDP90A) that measures the horizontal and vertical position of the light spot on the sensor surface, which is proportional to the deflection angles~$\phi_1$ and $(\phi_2 + \phi_2')$, respectively.
The schematics of the signal sensing method are shown in Figure~\ref{fig:sketch}~(a)~and~(b).
The vertical signal~$s_2$ mainly contains contributions from the main and pitch mode while the horizontal signal~$s_1$ is dominated by the yaw mode.
However, small cross-coupling contributions are possible due to non-ideal sensor alignment.
Instrumentation artifacts arise since the sensor response is non-linear containing spectral contributions at higher harmonics of the resonant modes, i.e. $nf_0$ and $nf_p$ for the fundamental and pitch mode, respectively, also see Figure~\ref{fig:asd}.
A separate measurement is performed to estimate the sensing noise contribution to the pendulum signal.
In this measurement, the path of the laser beam is altered such that it is reflected off a stationary mirror instead of the pendulum, thus containing no contribution from pendulum motion.

The seismically induced motion of the support platform is measured with a triaxial force-feedback seismometer (Nanometrics Trillium 120 QA) which outputs a signal~$w(t)$ proportional to the velocity.
The x- and y-axes of the seismometer measure the horizontal platform velocity perpendicular and parallel to the main pendulum motion, respectively, while the z-axis measures vertical velocity.

All signals are digitized with a data acquisition card with 14 bit resolution at 120 samples per second.

%
%
\section{ML models with and without witness data}
We consider the signals $\{s_i(t)\}$ over the look-back window $\kappa$ and predict the target $y(t)=s_j(t)$ over the look-ahead window $\tau$.
Witness channels $\{w_i(t)\}$ are included over the look-back and look-ahead window $\kappa+\tau$ corresponding to a scenario where the witness data is known ahead of time.
In Figure~\ref{fig:sketch}~(c) we illustrate this data layout.
The predicted evolution is therefore based on physical knowledge of the system from the past, deterministic instrumentation artifacts as well as witnessed noise.
Unwitnessed noise or experimental noise, such as quantization noise or flicker noise, may not be covered.

Two models are employed, one has access to the signals only (model~I), while the second one additionally has access to seismic witness channels (model~II).
For convenience we use the following notation for an interval of a signal $c_{a:b}(t)=c(t+a)$ where $t\in[0,b-a]$.
Our approach defines a multivariate multi-horizon time-series forecasting model \cite{shumway_time_2017} based on machine learning \cite{Lim_2021} to predict the target over the look-ahead time
\begin{align}
	\hat{y}^\text{I}_{t}(u)&=\mathcal{F}^\text{I}(s_{t-\kappa:t})(u) \\
	\hat{y}^\text{II}_{t}(u)&=\mathcal{F}^\text{II}(s_{t-\kappa:t}, w_{t-\kappa:t+\tau})(u)
\end{align}
where $u\in[0,\tau]$ is the look-ahead distance, $\mathcal{F}^{\text{I}, \text{II}}$ are artificial neural networks (ANN) with the associated network parameters $\theta^{\text{I}, \text{II}}$, $s_{t-\kappa:t}$ is the look-back signal and $w_{t-\kappa:t+\tau}$ are the witness channels.
The windows $\kappa, \tau$ are set such that multiple periods of the characteristic pendulum modes are resolved, encouraging the model to learn feature representations of those.
Due to the generic nature of ANNs, no further assumptions about the underlying model are required.
Intermediate feature representations are learned from a data-driven training procedure with the objective to infer network parameters corresponding to local minima on high dimensional loss landscapes.

Different network species have been demonstrated to be suitable for time-series forecasting, such as fully connected neural networks~\cite{Faraway_1998, Zhang_2003}, recurrent neural networks~\cite{Shih_2019, Han_2021, Shi_2022}, computational reservoirs~\cite{Tanaka_2019, Mandal_2022} or convolutional neural networks (CNN) \cite{Yang_2015, durairaj_convolutional_2022}.
In CNNs the layers are connected by convolutional operations with parametrized kernels of fixed size~\cite{Gu_2017}.
A kernel is defined by local space-invariant interconnections making the inner representations of the network equivariant to translations with respect to the prediction time~\cite{Zhang_1990, Goodfellow-et-al-2016}.
This allows for smooth translations of the prediction time, which is why we choose CNN as the main architectural components of $\mathcal{F}^{\text{I},\text{II}}$.
We set the activation of the output perceptrons to be hyperbolic tangent functions introducing non-linearity while capturing the oscillatory nature of the time-series.
In training, dropout regularization is employed encouraging the network to learn sparse representations and prevent overfitting~\cite{Srivastava_2014}.
Each channel of the multivariate input is represented by an isolated CNN sub-model.
Those sub-models are concatenated and post-processed by a sequence of fully connected layers having hyperbolic tangent activations as well, followed by the final layer having linear activation.
Further details are provided in App.~\ref{app:model-details}.

\begin{figure}
	\includegraphics{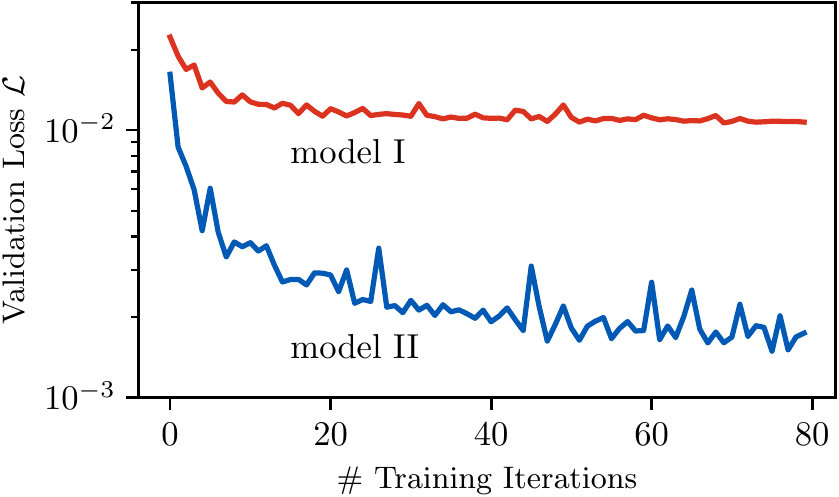}
	\caption{Validation loss over the number of training iterations. Model~I has access only to the look-back signal to form the target prediction and model~II additionally utilizes the seismic witness channels.}
	\label{fig:training}
\end{figure}

In this work we use the three seismic witness channels $w_x(t)$, $w_y(t)$ and $w_z(t)$ from the seismometer and for the look-back signal we employ the vertical $s_1(t)$ and horizontal $s_2(t)$ signals from the photodetector, as illustrated in Figure~\ref{fig:sketch}~(c).
We set the target to be the $s_2(t)$ signal and define the objective to be minimized during training, the loss function, to be the mean squared error of the predicted target
\begin{equation}
	\mathcal{L}^\text{I,II}=\frac{1}{\tau}\Vert \hat{y}_t^\text{I,II} -(s_2)_{t:t+\tau}\Vert^2.
	\label{eq:loss}
\end{equation}
Two disjoint datasets, the training-dataset and the validation-dataset~\cite{sohil_introduction_2022}, are sampled from the experiment such that the prediction times are uniformly distributed encouraging to learn equivariant feature representations with respect to shifts of the prediction time.
Based on the training-dataset, the associated network parameters are inferred as $\theta~=~\text{argmin}_\theta \mathcal{L}$ using stochastic gradient descent where the learning rate is dynamically adapted according to ADAM~\cite{kingma_adam_2017}.
On the other hand, the validation-dataset is used to define the validation-loss according to Eq.~\ref{eq:loss} allowing to quantify the training process.
We have reserved \SI{20}{\percent} of the overall record for validation.
For each training iteration, the datasets are resampled and served in batches. 
The validation-loss of both models converges as shown in Figure~\ref{fig:training}.
Model~I approaches $\mathcal{L}^\text{I}\approx 10^{-2}$ while model~II approaches $\mathcal{L}^\text{II}\approx 10^{-3}$ demonstrating that witnessed seismic noise improves the predictive capabilities.

Due to the multi-horizon forecasting over the look-ahead window $\tau$, there exist many predictions $\hat{y}_{t-u}(u)$ at time $t$ corresponding to different look-ahead distances $u$.
This motivates the definition of the prediction as the weighted average
\begin{equation}
	\tilde{y}^\text{I,II}(t) = \int_{0}^\tau \mathrm{d}u\,x(u)\hat{y}^\text{I,II}_{t-u}(u),
	\label{eq:prediction}
\end{equation}
where $x(u)$ is a normalized weight function on $[0,~\tau]$.
For the upcoming discussions we choose the weight function to be uniform $x(u)=1/\tau$ so that contributions near the prediction time $t$ as well as predictions far into the future $t+\tau$ are weighted equally.
The target can be written as
\begin{equation}
	s_2(t) = \tilde{y}^\text{I,II}(t) + r^\text{I,II}(t),
	\label{eq:reduced}
\end{equation}
where $r^\text{I,II}(t)$ is the noise-reduced signal containing unpredicted contributions.

%
%
\section{Results of model noise reductions}
\begin{figure}
	\includegraphics{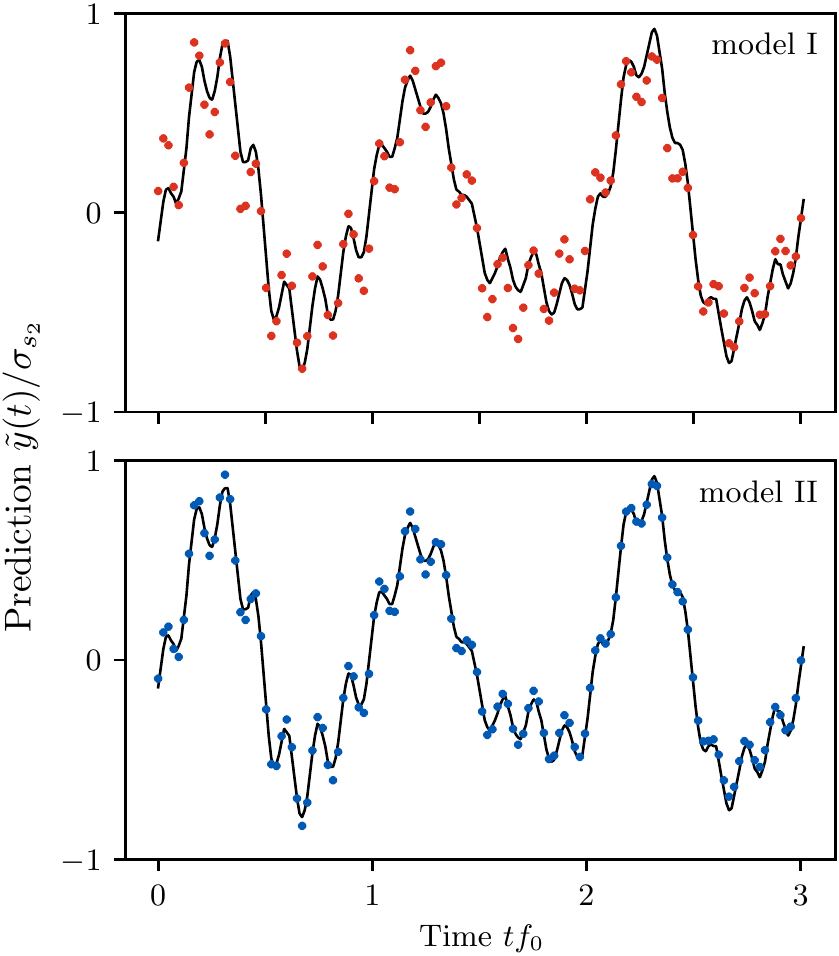}
	\caption{
		Example predictions (dots) from the model without (I) and with seismic witness channel (II).
		The target $s_2(t)$ is depicted as a black continuous line for visual simplicity.
		The y-axis is in units of the standard deviations $\sigma_\text{s2}$.
	}
	\label{fig:example_predictions}
\end{figure}
The predictions $\tilde{y}$ according to Eq.~\ref{eq:prediction} were evaluated over the validation dataset.
In Figure~\ref{fig:example_predictions} we show a single prediction sample, where the target~$s_2(t)$ is shown as well, for comparison.
The predictions of both models contain periodic structures close to the expected target suggesting that the model synthesizes the phase space initial conditions from the look-back signal, allowing to integrate the inferred dynamics to obtain the state space evolution.
Due to the harmonic nature of the dynamics we conclude that the underlying CNN acts as a Fourier transformation synthesizing the amplitudes and phases of the modes from the unperturbed motion of the pendulum.
This works especially well as the network identifies a discrete number of features corresponding to sharp peaks in the spectrum due to the high-Q factor.
Seismic witness data qualitatively improves the predictive capabilities suggesting that the pendulum motion is correlated with the seismic noise.

\begin{figure*}
	\includegraphics{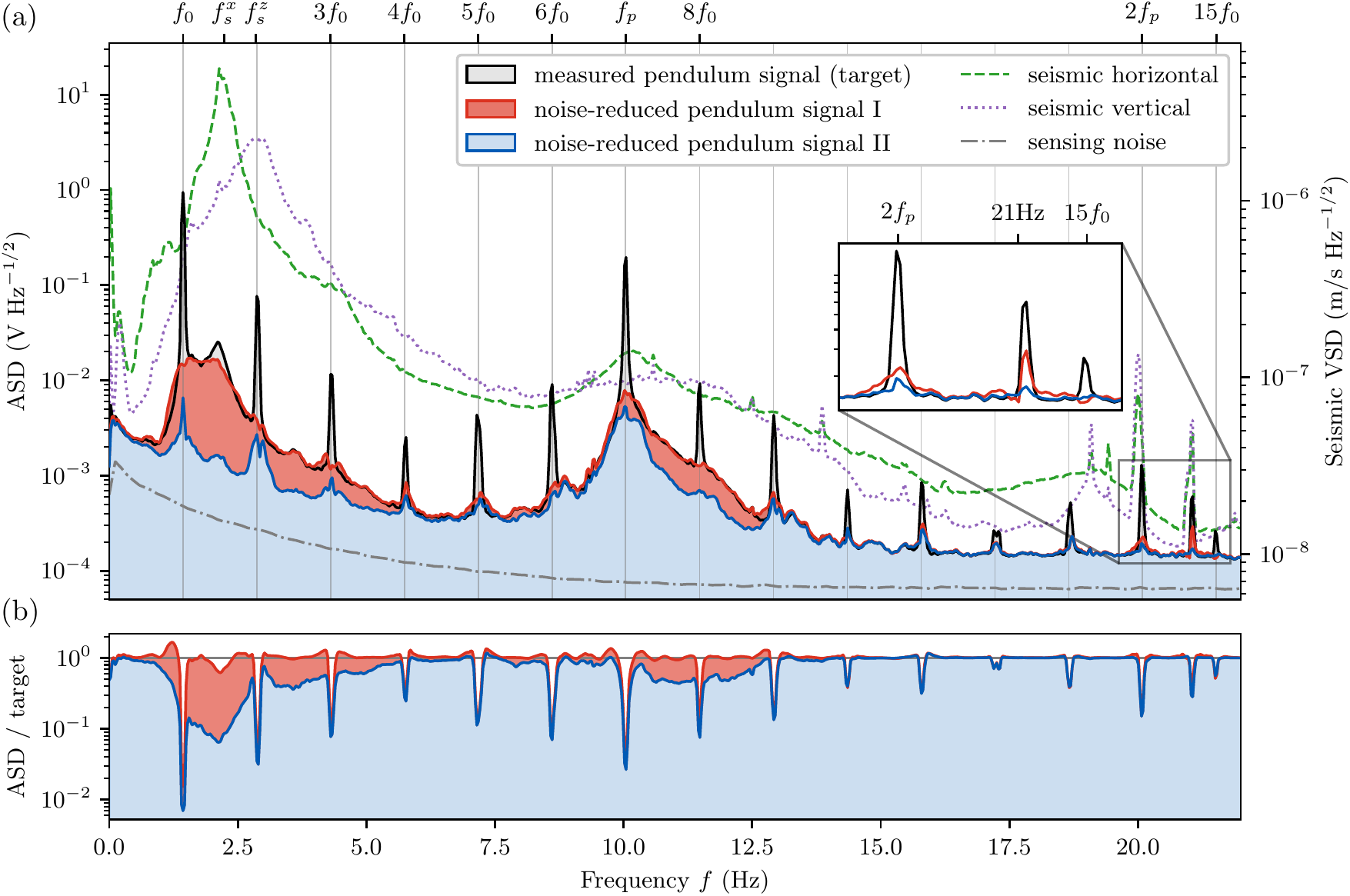}
	\caption{
		\textbf{(a)} Amplitude spectral density (ASD) of the measured and noise-reduced pendulum motion (top of colored areas) as well as the velocity spectral density (VSD) of the seismic witness sensors (dashed and dotted lines).
		The noise-reduced ASDs $r^\text{I,II}(f)$ are taken from the difference between the measured pendulum signal $s_2$ (target) and the predictions $\tilde{y}^\text{I,II}$ of each model, see Eq.~\ref{eq:reduced}.
		Both models are able to reduce periodic noise at the resonances~$n f_0$.
		However, the model utilizing seismic witness channels (noise-reduced pendulum signal II) allows for a more broadband noise reduction close to the sensing noise level (dash-dotted line).	
		The inset plot depicts high frequency noise reduction where we want to highlight the off-resonant noise reduction near the seismic peak at \SI{21}{Hz}.
		\textbf{(b)} Relative ASD of the noise-reduced signals compared to the target signal.
		Resonant features are reduced by up to two orders of magnitude by both models, while off-resonant noise is reduced by up to one order of magnitude with model~II utilizing seismic witness channels.
	}
	\label{fig:asd}
\end{figure*}
Next, we show the spectral densities of the pendulum and seismometer signals using Welch's method~\cite{Welch_1967} over samples of size $2^{12}$.
Figure~\ref{fig:asd} shows the amplitude spectral density (ASD) of the target $s_2(f)$ and the reduced signals $r^{\text{I},\text{II}}(f) = (s_2 - \tilde{y}^{\text{I},\text{II}})(f)$ from both models according to Eq.~\ref{eq:reduced}.
Also shown is the velocity spectral density (VSD) of the observed seismic perturbations, where the maximum is near the main pendulum mode~$f_0$.
The ASD of the target $s_2(f)$ shows pronounced peaks at the resonances $f_0$ and $f_p$.
The peaks at multiples of $f_0$ and $f_p$ correspond to the non-linearity of the photodetector, see Section II, as expected for sinusoidal signals passed through a non-linear element~\cite{moir_rudiments_2022}.

To compare the results that we achieve with our machine-learning based method, with an established noise reduction method, we apply linear Wiener filtering to the same data. As we discuss in App.~\ref{app:linear}, we find only a limited success of linear noise filtering, because only linear correlations between the witness channel and the target are included~\cite{vega_rapid_2013}.

Both our models, on the other hand, resolve those non-linear artifacts well as evident by the dips in the ratio of the ASD and the target spectral density, displayed as ASD/target in Fig. 4 (b), at all harmonics of $f_0$ and $f_p$.
Such monochromatic features are well extrapolated solely from the signal look-back, as on the timescale of the prediction window, their phases and amplitudes are only affected by resonant seismic transients, which occur rarely.
Off-resonant excitations, however, are time-local and hence not predictable from the signal look-back.
Here, the advantage of using the seismic witness channels becomes evident. 
In the spectral region of high seismic activity (\SIrange{0.5}{5}{Hz}), model~I could reduce the root mean square amplitude of the signal by $71\%$, while utilizing witness channels, model~II further lowered the amplitude by a factor of 4.

In the frequency region around the pitch mode (\SIrange{7}{13}{Hz}), there is an asymmetry in the spectral density of the target.
Left of the pitch mode resonance~$f_p$, the target ASD is lower than on the right although the spectral density of the seismic background is roughly the same on both sides.
Also, on the left side, model~II delivered no significant improvement over model~I, while on the right side around \SI{11}{Hz}, witness data allows to reduce the off-resonant ASD by a factor of 4.
We assume this asymmetry is caused by an interaction between the main- and pitch-mode resonances as horizontal suspension point movement has counteracting effects on the rotation of the test mass in the two modes.
This would make the reflection angle measurement insensitive to modal excitations at some frequencies.

Near \SI{21}{Hz}, a peak appears both in the ASD and seismic VSD.
This peak could be partially reduced by model~I and removed almost completely by model~II, showing that high frequency features are resolved as well.

At most frequencies, the noise-reduced spectrum of model~II follows the sensing noise closely, indicating that most of the witnessed and predictable noise has been subtracted from the target.
The sensing noise level is dominated by quantization noise of the data acquisition system at high frequencies, where it follows a flat line.
Towards low frequencies, it shows an increase, likely due to flicker noise in the detector amplification circuits~\cite{horowitz_art_2015}.

Last we want to discuss the role of the weight function~$x(u)$.
Without any further assumption we have estimated the parameters where model~I shows the best results when only the immediate target prediction $\tilde{y}_{t}^\text{I}(0)$ is considered, i.e. $x^\text{I}(u)=\delta(u)$, while for model~II, the weights decay exponentially $x^\text{II}(u)\sim~e^{-au/\tau}$.
Therefore, witness data allows to predict subsequent future states.
However, finding optimal weights depends on the specific physical application and further assumptions must be made.
For example, witness data and signals might encounter phase differences due to instrumental delay or spatial separation between the measurement devices.
Feed forward control also imposes model dependent requirements to the weights.
Manufacturing optimal weights is therefore a rich and significant task depending on the application's design.

%
%
\section{Conclusion}
In conclusion, we have presented a machine learning based time-series forecasting model to predict the seismically excited oscillation of a pendulum having a resonance frequency of \SI{1.4}{Hz} and a Q factor of $6 \cdot 10^4$ with the addition of a three-directional seismic witness sensor.
The spectral analysis of the pendulum motion reveals that without the witness channels machine learning can well predict the displacement amplitudes related to the pendulum resonance.
The pendulum motion at off-resonance frequencies is dominated by the continuous changes of the seismic field.
As expected, the amplitudes at these frequencies could only be predicted when the model had access to the information in the seismic witness channels.

We conclude that our trained neural network learned both the natural behavior of the pendulum and the transfer function from the witnessed seismic noise to the displacement of the pendulum suspended mass, including instrumental artifacts such as non-linearities of the sensor used to measure the pendulum oscillation.
Our approach enables flexible multivariate sensor layouts as the model learns the correlations in a model-free approach and no direct measurements of transfer functions have to be performed.
The high predictability of the pendulum motion demonstrates the applicability of machine learning for feed-forward suspension control to counteract pendulum excitation through the local seismic field.

\section{Outlook towards corrective feed-forward}

In gravitational-wave detectors, the control of pendulum suspensions is a complex effort to strike a balance between stabilization and minimization of introduced noise, requiring advanced control strategies~\cite{strain_damping_2012,aston_update_2012,hartwig_mechanical_2022}.
The machine learning approach in combination with external sensors, as demonstrated here, can be used in the future to reduce the seismic excited motion of a pendulum in advance. A feed-forward control loop would exploit the knowledge of the transfer function in question to correctively stabilize, for example, the pendulum's suspension point against incoming disturbances. Alternatively, it could act on the passively isolated platform on which the pendulum suspension is constructed.
Or, this control loop would in our proof-of-principle experiment stabilize the plate of our optical table.
In all cases, there is the significant advantage that the control loops at the end of the chain, which act directly on the position of the mass of the pendulum, have to do so with less force. It is believed that lower forces on the pendulum will lead to a reduction in the rate of non-Gaussian transients, so-called glitches~\cite{Derosa_2012}.

Conventional active stabilisation is usually implemented with linear control systems, where the control forces acting on the system are modelled by transfer functions in frequency space~\cite{Astrom2009}. Active stabilisation using a machine learning approach in combination with external sensors has the advantage that non-linear disturbances can also be corrected.
Alternatively, adaptive filtering techniques allow the forward transfer function to be optimised during operation~\cite{driggers_active_2012}. In principle, even Newtonian noise can be cancelled if mass displacements that cause gravity fluctuations are observed by additional sensors~\cite{harms_terrestrial_2019}. 
Our work provides the proof-of-principle that machine learning can be used to predict the motion of systems coupled to an environment. This supports the idea that machine learning-based corrective forward stabilisation is a promising platform for improving the signal-to-noise ratio in next-generation gravitational-wave detectors.

%
%
\section{Acknowledgments}
We thank Lukas Broers and Jim Skulte for very helpful discussions.
This work is funded by the Deutsche Forschungsgemeinschaft (DFG, German Research Foundation) - SFB-925 - project 170620586, the Cluster of Excellence 'Advanced Imaging of Matter' (EXC 2056) project 390715994 and the Cluster of Excellence 'Quantum Universe' (EXC 2121) project 390833306.

\bibliographystyle{apsrev4-1}
\bibliography{ml-interferometer-arXiv.bib}

%
%
\appendix
\section{Model Details}
\label{app:model-details}
\begin{figure*}
	\includegraphics[width=17cm]{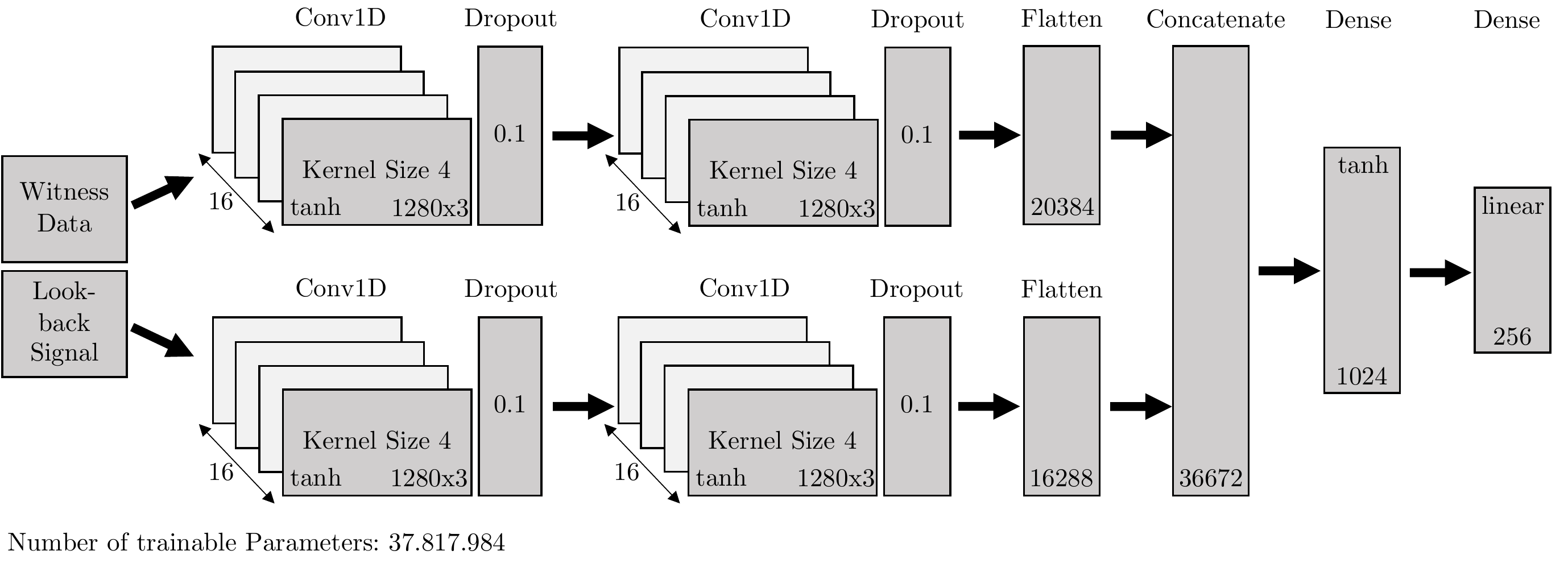}
	\caption{The ANN used in this work. The look-back signal and witness data is processed in two separate CNN sub-models. The subsequent outputs are concatenated and processed by a sequence of dense layers.}
	\label{fig:nn}
\end{figure*}
In this section we outline details of the artificial neural network (ANN) model used in this work.
Let's denote the discretized temporal $n$-dimensional signals as $c_{t_1:t_2}\in\mathbb{R}^{\lfloor f(t_2-t_1)\rfloor\times n}$ with $t_1 < t_2$ and the sampling frequency $f$ of $c$.
The multi-horizon time-series target prediction of the look-back signal $s_{t-\kappa:t}$ and witness channels $\{w_{t-\kappa_w:t+\tau_w}\}$ over the look-ahead window of size $n_y$ is
\begin{equation}
	\hat{y}_{t} = \mathcal{F}(s_{t-\kappa_s:t}, \{w_{t-\kappa_w:t+\tau_w}\}) \in\mathbb{R}^{n_y}.
\end{equation}
where $\mathcal{F}$ is an ANN.
In our specific case we have $s=(s_1(t), s_2(t))$ as the two dimensional pendulum signal and $w=(w_x(t), w_y(t), w_z(t))$ the three-directional seismic witness signal.
All signals share the same sampling frequency $f=\SI{120}{\hertz}$ and the target prediction is
\begin{equation}
	\hat{y}_{t} = \mathcal{F}^\text{II}(s_{t-\kappa:t}, w_{t-\kappa:t+\tau}) \in\mathbb{R}^{n_\tau}.
\end{equation}
We set $n_\kappa=\lfloor f\kappa\rfloor=2^{10}$ and $n_\tau=2^{8}$.
The prediction target is $y_t=(s_2)_{t:t+\tau}$.
Temporal signals are assumed to be stationary as the pendulum is contained within a vacuum chamber and hence under stable environmental conditions (such as pressure and temperature).
Seismic noise is assumed to be stationary on time scales of the experimental run.
Therefore, the temporal signals are standardized as $s_{1,2}, w_{x,y,z} \sim \mathcal{N}(0,1)$ in preprocessing.

The input space is $\mathcal{X}^\text{II} = \mathbb{R}^{n_\kappa\times2}\times \mathbb{R}^{(n_\kappa+n_\tau)\times3}$ with the corresponding labels $\mathcal{Y}^\text{II} = \mathbb{R}^{n_\tau}$.
The isolated model has $\mathcal{X}^\text{I} = \mathbb{R}^{n_\kappa\times2}$ using the same labels $\mathcal{Y}^\text{I} = \mathcal{Y}^\text{II}$.
The datasets are sampled from the uniform distribution $U(I)$ of prediction times on the sampling interval $I$
\begin{align}
	T_I^\text{I} &= \{((s_{t-\kappa:t}), y_t)|t\sim U(I)\}\\
	T_I^\text{II} &= \{((s_{t-\kappa:t}, w_{t-\kappa:t+\tau}), y_t)|t\sim U(I)\}
\end{align}
The training-dataset $T^\text{I,II}_{[0,qt_\text{max}]}$ and validation-dataset $T^\text{I,II}_{[qt_\text{max},t_\text{max}]}$ are disjoint where we have $t_\text{max}=60\text{h}$ and set $q=0.8$.
After each training iteration, training and validation data is resampled.
The network parameters $\theta$ are inferred over a training of 500 training iterations with a batch size of 128 using a training dataset of size $10^5$.
We introduce the prediction at time $t$ as the weighted sum of all target predictions containing that particular time
\begin{equation}
	\tilde{y}_{t} = \sum_{j < n_\tau}X_j\cdot(\hat{y}_{t-j})_{j}
	\label{eq:h_discrete}
\end{equation}
where $X_j$ are normalized weights.

$\mathcal{F}^\text{I,II}$ was implemented using Tensorflow \cite{Abadi_2016}.
Witness channel data $(n_\kappa + n_\tau,3)$ is the input of two sequential 1D convolutional layers of depth 16 and kernel size 4 having dropout of 0.1.
A similar convolution is applied to the look-back signal $(n_\tau,2)$.
The output from the sub-models are concatenated and processed thru a dense layer of size $4n_\tau$ having tanh activation followed by a linear dense layer producing the target prediction of size $n_\tau$.
In Figure~\ref{fig:nn} we provide a visual representation of $\mathcal{F}^\text{II}$.
The network was trained on a single compute node providing $64\text{GB}$ of RAM.
For our concrete model, the number of trainable parameters scales linearly in the number of sub-models, that is approximately $2 \times 10^7$ trainable parameters per sub-model.
Therefore, large sensing arrays can be implemented on compute clouds providing distributed inference.

\section{Linear Filter}
\label{app:linear}
In this section, we construct a forward linear prediction Wiener filter~\cite{vega_rapid_2013} and compare it to the proposed non-linear model.
We construct a linear model to predict the target $s_2(t+\tau)$ at a look-ahead time of $\tau$ utilizing history and witness data over look-back time of $\kappa=0.5\si{\second}$.
The loss function is given as the mean squared error
\begin{equation}
	\mathcal{L}_\tau = \mathbb{E}\left[||s_2(t_{i}+\tau) - \hat{y}(t_{i}+\tau)||^2\right],
	\label{eq:L-wiener}
\end{equation}
with the prediction
\begin{equation}
	\hat{y}(t_{i}+\tau)=\sum_{j=0} h^s_j s(t_{i-j}) - \sum_{j=0} h^w_j w(t_{i-j}+\tau),
\end{equation}
where $h^s \in \mathbb{R}^{60\times 2}$, $h^w \in \mathbb{R}^{60\times 3}$ are finite impulse response filters of size $60$, and $s(t)\in \mathbb{R}^2$ and $w(t)\in\mathbb{R}^3$ are the respective signals.
The witness channels $w(t)$ and look-back signal $s(t)$ are as given as in the latter section.
The filters $h^j$ are inferred via stochastic gradient descent over the same dataset and training configuration as used for the non-linear model.
The residual ASD $(s_2 - \hat{y})(f)$, the target ASD $s_2(f)$ and the VSD $w_i(f)$ are calculated using Welch's method~\cite{Welch_1967} over samples of size $2^{12}$.

\begin{figure*}
	\includegraphics{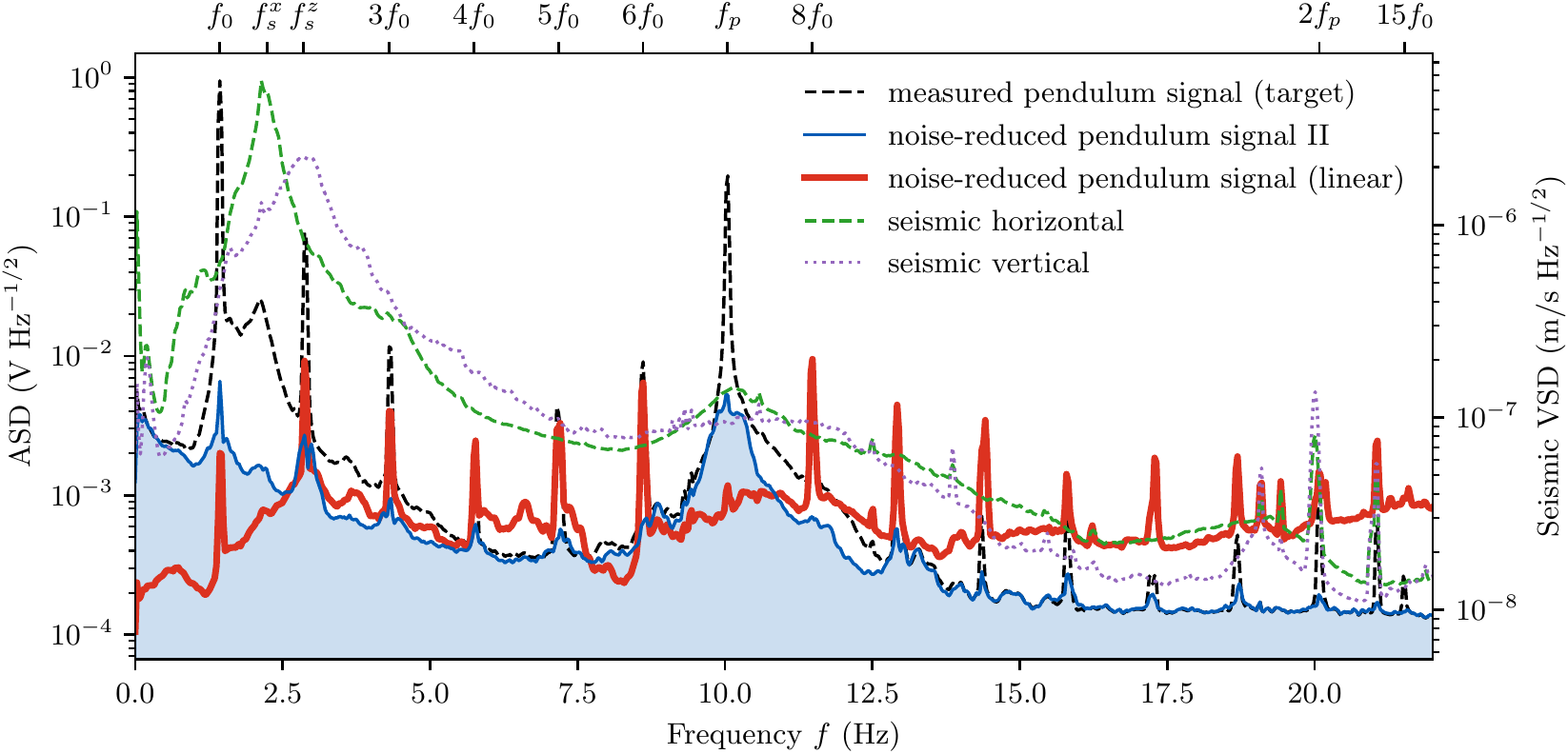}
	\caption{Residual ASD of the linear Wiener filter for the prediction target $s_2(t+1/f)$ (red line), that is the prediction of the immediate following sample corresponding to the sampling frequency of $f=\SI{120}{\hertz}$.
	The target signal is shown as a black-dashed line and the noise-reduced pendulum signal from the non-linear machine learning model II is shown as a blue line.
	The horizontal (vertical) VSD of the seismometer is depicted as a green-dashed (purple-dotted) line.}
	\label{fig:linear-1}
\end{figure*}

\begin{figure*}
	\includegraphics{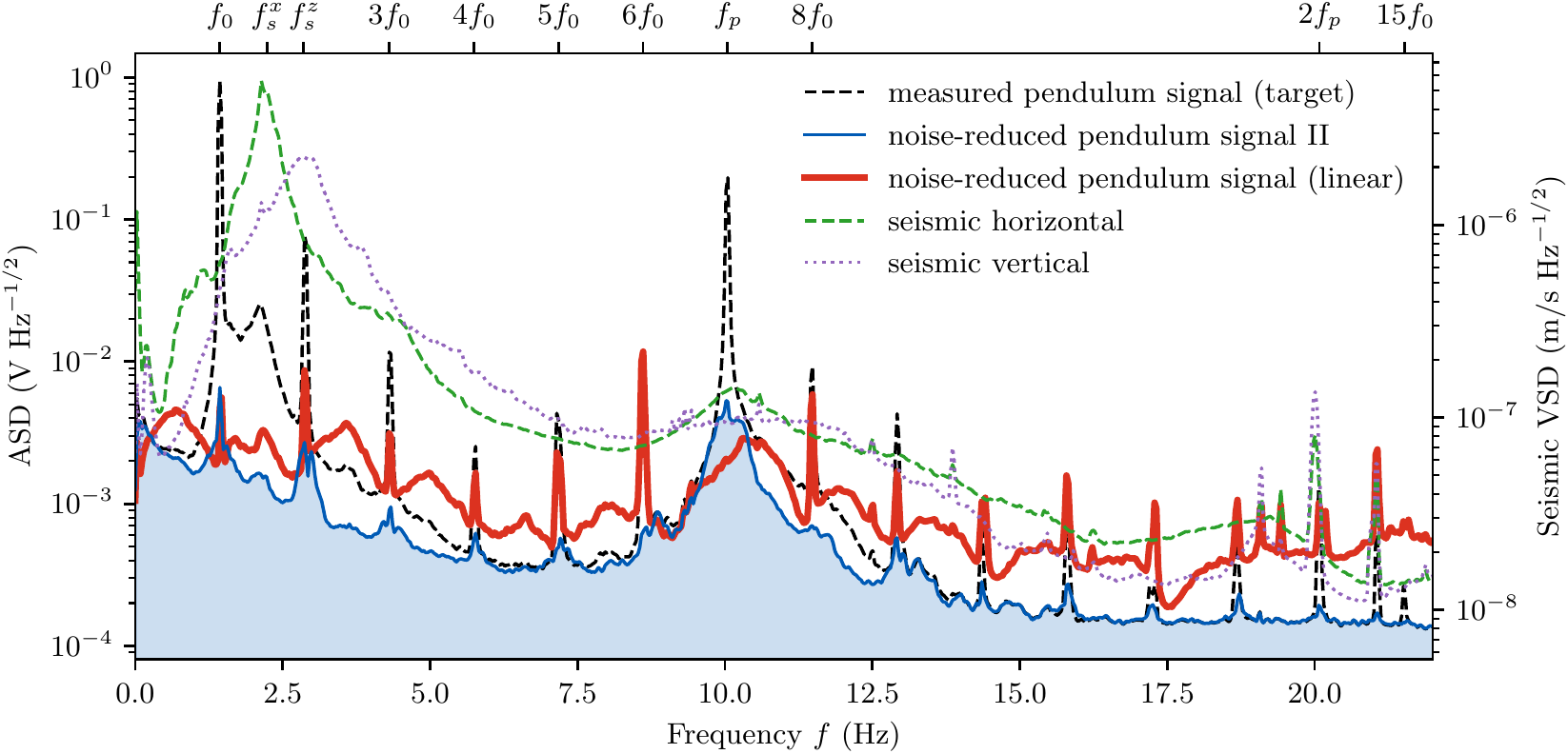}
	\caption{Residual ASD of the linear Wiener filter for the prediction target $s_2(t+\SI{0.2}{\second})$ (red line). The look-ahead time of the linear Wiener filter of \SI{0.2}{\second} is small compared to the look-ahead time of the non-linear model, that is \SI{2}{\second}.
	The target signal is shown as a black-dashed line and the noise-reduced pendulum signal from the non-linear machine learning model II is shown as a blue line.
	The horizontal (vertical) VSD of the seismometer is depicted as a green-dashed (purple-dotted) line.}
	\label{fig:linear-24}
\end{figure*}

We consider two examples.
In the first example, the look-ahead time is set to the immediate following sample, i.e. $\tau=1/f$. In the second example, the look-ahead time is set to $\tau=\SI{0.2}{s}$, corresponding to 24 samples.
Note that for the proposed non-linear model we use a look-ahead time of $\tau\approx \SI{2}{s}$.
Figure~\ref{fig:linear-1} shows the ASD of the noise-reduced signal at a look-ahead time of $\tau=1/f$.
The signal can be extrapolated easily in the low frequency spectra.
Here, the linear filter reduces the ASD of the signal by $95\%$ for \SIrange{0.5}{5}{Hz}, while the non-linear model reaches values of $92\%$.
For \SIrange{5}{8}{\hertz} and $f > \SI{13}{\hertz}$, the noise-reduced signal is actually enhanced compared to the target.
Here, white-noise associated with the finite impulse response filter is greater than the ASD of the target.
At the main pendulum mode $f_0$ and the pitch mode $f_p$, the ASD is reduced by several orders of magnitude.
As expected, the higher harmonics $n f_0$ corresponding to non-linear instrumentation artifacts are not captured by the linear filter.
For look-ahead times of $\tau=\SI{0.2}{s}$, the overall ASD is less reduced, as shown in Figure~\ref{fig:linear-24}.
Here, the linear filter reduces the ASD of the signal by $86\%$ for \SIrange{0.5}{5}{Hz}.
Near the pitch mode, the reduction is less pronounced as in the latter case.
The level of white noise is similar in both cases, as it depends on the filter size~\cite{vega_rapid_2013}.
Hence, at high frequencies of $f > \SI{15}{\hertz}$, the ASD are similar for both examples.
We see higher harmonics exceeding white-noise levels and find that other resonances in the seismic VSD induce additional peaks above the target ASD.

Our proposed model provides crucial improvements compared to the discussed linear model.
The residual ASD is lower than the target ASD over the whole frequency domain, which is not the case in the linear approach, due to white-noise induced by the finite impulse response filters.
Non-linear instrumentation artifacts are captured by the non-linear model, in contrast to the linear model.
Our proposed model outperforms the predictive capabilities of the linear model as the linear model shows a less pronounced reduction of the ASD at only $10\%$ of the look-ahead time $\tau$ used by the non-linear model.
We note that we have employed a low-complexity optimization approach to infer the filter weights $h$.
The success of this approach depends on careful tuning of the step-size used within the stochastic gradient descent method.
We emphasize that a more sophisticated linear filter may be more efficient in the regard of noise-reduction.

\end{document}